\documentstyle[12pt]{article}

\title{\huge Phenotypical Behavior and Evolutionary Slavery}

\author{\huge Andr\'e C. R. Martins\\
\\
\it Banco Central do Brasil, Studies and Research Department\\
\it av. Paulista, 1804, $4^{o.}$ andar, S~ao Paulo - SP, Brasil\\}
\date{}

\input tcilatex

\begin{document}

\maketitle

\begin{abstract}
The Prisoner Dilemma has basically two known type of evolutive answers that allow 
cooperation among individuals. One proposes cooperation is possible among close 
relatives.  The other is a strategy on when to cooperate and when not to, according 
to the actions of the other players. An example of these strategies is playing 
tit-for-tat. This paper proposes a third and completely different solution to the 
evolutionary problem of cooperation, based on the fact that a specific gene needs 
not fix completely the behavior of the individual. It is a valid genetic instruction 
to create individuals with the same gene that have a chance of behaving one way and 
a chance of behaving differently, or, in other words, have different phenotypical 
strategies. We will see that, depending on the paremetrs of the game, there is a 
better genetic strategy than the previous ones mentioned above. That is done creating 
individuals with a phenotypical behavior to serve the other members of their species 
and others who will be served. This creates a type of evolutionary slavery. The 
species is divided into two kind of beings, called here as leaders and  servants, and 
we show that this is evolutionary favorable to the species, as well as a stable 
solution. Moreover, genes playing always cooperate inside the family or even 
tit-for-tat are found to have no barrier against an invasion by this new strategy. It 
is also shown that this result does not apply only to the Prisoner Dilemma but can be 
generalized to other evolutionary games among individuals, as long as they have the 
right parameters, actually increasing the fitness of the species.  Depending on the 
parameters of the game it is a stable and best solution to choose non-dominant 
strategies, different for the leaders and the servants, even when a dominant strategy 
is available to them. Possible applications of this strategy to problems as the 
appearance of multi-cellular beings and some charactstics of human behavior are 
proposed and shortly disccussed and it is verified that this new approach to the 
solution of Evolutionary Game Theory problems can lead to some very interesting 
consequences and explanations of known facts.
\end{abstract}

\section{Introduction}

A problem that has always puzzled evolutionary game theorists is the amount
of observed cooperation among individuals from the same specie or even
belonging to different species, even when it would be harmful for each
individual to cooperate. Animals warn their companions about the approach of
a predator shouting to their comrades, attracting attention to themselves and,
therefore, increasing the chances that the predator will notice them and
choose them as the next meal. Animals seem to help their fellows more often
than it would be expected in these situations and this problem has been
offered two solutions. The problem here is that, even if cooperation is the
best solution for all the individuals, it is not stable in a Prisoner
Dilemma circumstance. Any mutant that would decide not to cooperate would
see its success increasing, as the other animals help him, but he doesn't
run the risks of attracting the predators to himself by not shouting when it
could.

However, the structure that really is playing these games and "learning" a
way to improve its strategy, is not the individuals, who are mortal and,
no matter how successful they become, will disappear. It is their genetic
code that is continuously changing itself, to adapt to new circumstances or
simply because it has found out a better way to do things. This idea was 
first proposed by Richard Dawkins~\cite{dawkins}.

Hamilton~\cite{hamilton} proposed that, for a rare gene to survive, it would
make sense to cooperate with our brothers, as they would have a 50\% chance
of having that same gene. The percentage would go down fast to $1/8$ of
chance between first cousins, so the cooperation would be something that
could happen inside families, among close relatives. Our own survival would
still be more important than that of our relatives, but we could warn them
about danger, as long as that would make their survival, along with our own
genes, much more likely. In extreme cases, like in an ant colony where all
ants share exactly the same genetic code, the advantages to a gene survival
gained from cooperation would be even stronger and the behavior of soldier
ants, who never reproduce and die without hesitation to protect their colony
would make even more sense.

Axelrod\cite{axelrod} suggested the second solution to the dilemma in the
form of a strategy. He created several programs who would compete inside a
computer, in an environment where the Prisoner Dilemma were to happen and
gave it of them a strategy to try to defeat the other competing programs. He
found out that the program that cooperated with programs that had previously
cooperated with it and did not when in the other case would systematically
win over the non-cooperating strategies. The reason for that was quite
simple. When facing a non-cooperative algorithm, our algorithm would not
risk itself and would not cooperate getting the best possible return.
However, at least when facing copies of itself, both would cooperate,
allowing for a best overall performance of the individuals.

Therefore, the theoretical possibilities for cooperation, so far, seemed to be 
some limited true cooperation among members of a family and the possibility of 
adoption of an strategy that would take us to be cooperative only when the
individual we are interacting with has not failed to cooperate with us in the
past. In this paper, I propose a third possible source of cooperation among
individuals of a species. The key to such a cooperation is in the way that 
genes do influence the real behavior of one individual. So far, it has been
considered in the literature that having a specific genetic code forces you 
to adopt a specific strategy~\cite{game1,game2}. 

But this doesn't have to be that way. A mutant gene could come up 
with instructions that a percentage of the individuals possessing it would 
behave in a way, while others would do it differently. There might be some 
sort 
of trigger determining which individuals will follow which strategy, like the
position of the individual in some hierarchy or anything as trivial as the
fact of been born during a cold or a warm day. What makes the mutation here
proposed a possible winner is not how this decision is taken, although for
specific problems there might be best answers, but the fact that some 
individuals with exactly the same code will, at some point, decide to act
one way or the other.

This possibility for different behaviors associated with the same gene, or the 
appearance of phenotypical properties nor dictated by the gene will, as we 
will see, open the possibility for the appearance of individuals who exist 
merely 
to serve their peers, while the others have a more comfortable life. Therefore, 
the use of the name evolutionary slavery in the title of this paper.

\section{The Prisoner's Dilemma and its Strategies}

The Prisoner Dilemma happens in a very simple game between two players. In 
the
example above, where an individual must decide whether or not warn its 
relative
about the arrival of a predator, possibly calling the attention of the hunter 
to
itself, let's say that, in average it has a chance of 10\% of being killed if
she and her companion cooperate, 50\% if none of them does and 90\% of chance 
of
been killed to the individual who cooperates when its partner does not, while 
the
non-cooperative specimen will always stay safe if her partner cooperates. 
This
can be represented in a matrix form as bellow:

\begin{equation} \label{game}
\left ( \begin{array}{cc}
		-0,5 & 0\\
		-0,9 & -0,1

\end{array}
\right )
\end{equation}

There are some known good strategies, in the sense that your decision doesn't 
have to be 
never cooperate. As we mentioned those strategies are cooperate only with 
your
very close relatives or cooperate with everybody but those who have failed to
cooperate with you before. Both are good answers to the Dilemma and they can 
also 
be used together, so that you would always cooperate with someone who is a 
very 
close
relative, regardless of their past actions, and cooperate with your friends 
as 
long as they cooperate with you. It is quite possible that we won't find 
other 
fixed strategies that are not variations on these two themes and that work 
well
in an evolutionary dynamics, not leading to extinction. However, a small 
change
in the way we look at the problem immediately opens a new possibility. 

So far, we have always considered that when an individual has a specific 
genetic
code, the strategy he will choose is already completely determined by that 
code,
like it was set in stone. In other words, we are imposing that the strategy 
one
adopts had a purely genetical component, leaving nothing to the phenotype. 
Biologists know very well that same genes can lead to different external 
manifestations, different phenotypes. Therefore, there is really no reason 
why we
should assume that for a specific gene, we should have a specific strategy. 
As we
will see, a mutant gene that allows for different strategies associated with 
it
could also flourish in a world where everybody else never cooperates or even 
in
situations where everybody else is playing tit-for-tat.

Now imagine that our species is well adapted to its environment, meaning that 
it 
uses a strategy as good as everyone else. It has means to identify its close 
relatives reasonably well and the genes make it sure that cooperation happens 
in 
this case. Tit-for-tat may or may not be used by everybody; if it is, just 
assume 
our species does it as well. Now, a mutation occurs and we have that the new 
gene
determines that part of the individuals who have it behave in a certain way, 
and 
the rest of the population
behaves according to different rules. We are not concerned at this point with 
what 
triggers the individual decision about which behavior manifests in each case. 
We
will return to this point later. 

The two possible phenotypes for that gene differ in the strategy they assume 
when 
facing someone who will probably have the same genetic code (a close 
relative,
something the species already knows how to recognize). When meeting 
strangers, they 
still act the same way, using the well-tested strategy their species 
developed.
In the internal relations that lead to the Prisoner Dilemma, part of the 
population, that
I will name as leaders, for reasons that will become clear very soon, 
never
cooperates, while the other part, the servants, cooperate everytime. This way, 
a
leader will always get the maximum benefit when interacting with a servant, 
while
the servant has to support the burden of the worst result in return. The only 
advantage for a servant is the fact that the other servants will always
cooperate with him.

For the gene, if the proportion of leaders and servants is right, the 
advantage
obtained by the leaders is far stronger than the problems caused by their 
non-cooperative attitude. Against a population who never cooperates, this 
strategy 
may even improve the fitness of the servants, for the right parameters, as what 
they
have to lose from the non-cooperation of the leaders may be compensated by the
cooperation from other servants. The reason for that is quite easy to see. 
Suppose 
that half of the population of the mutant gene is consisted of leaders and the 
other half of servants. The interaction with individuals who do not have the 
gene
is not altered and everybody always get a $-0.5$ in average, just as the 
non-mutants get every time. Among themselves, the servants get a $0.1$ half of 
the
time and a $-0.9$ the other half, for the same $-0,9$ average. If the penalty 
for
cooperating with a non-cooperator was just a little smaller or if the result 
obtained for two non-cooperating individuals a little worse, the servant 
population 
would, in average, have an evolutionary advantage when compared with the 
non-mutant
gene. 

In any case, it is easy to see that the non-cooperative population has no 
protection at all against an invasion by this genetic strategy. For a 
percentage of the total
population $\epsilon$ having the mutant gene, the average result for the 
non-mutants
doesn't change from $-0.5$. The mutants, however, get a

\begin{equation}
-0.5 +\frac{\epsilon^2}{8}
\end{equation}

\noindent
result, with is always better than what the 
non-mutant 
population gets, for every value of $\epsilon$.

Let's see how this works on the general case where we have an invasion by
a percentage of $\epsilon$ mutants, where there is a probability of $p$
for a specific mutant to be a leader (and, of course $1-p$ that he will be
a servant). The Prisoner Dilemma takes here the general form

\begin{equation}\label{prisoner}
\left ( \begin{array}{cc}
		a & c\\
		d & b
\end{array}
\right )
\end{equation}

\noindent
where $c>b>a>d$. That is, if you don't cooperate and your opponent does, you 
get $c$ and he $d$, if both cooperate, both get $b$ and if none does, both
get $a$.

In this case, we will have that a non-cooperative population would get as
a result of its actions always $a$, no matter if they are interacting with 
mutants or with non-mutants. The case for the mutants, however has to be 
divided in two parts, the gains obtained by the leaders $G_l$ and the gain 
for servants $G_s$. These are given by

\begin{eqnarray}\label{gain1}
G_{l} & = & \left ( 1-\epsilon \right ) a + \epsilon \left [ pa+\left ( 
1-p \right) c \right ] \nonumber \\
G_{s} & = & \left ( 1-\epsilon \right ) a + \epsilon \left [ pd+\left ( 
1-p \right) b \right ] \nonumber
\end{eqnarray}

The average gain $G$ for the gene would then be given by

\begin{equation}
G = p G_{l} + \left ( 1-p \right ) G_{s} = a \left ( 1-\epsilon \right ) +
\epsilon S \nonumber
\end{equation}
\noindent
where

\begin{equation}
S= \left [ p^2 \left ( a+b-c-d \right ) + p \left ( c+d-2b \right )
+b \right ] \nonumber
\end{equation}

If we have no leaders ($p=0$), we have that $G=a( 1-\epsilon ) + b
\epsilon$. That is clearly better than the result for non-mutants, $a$,
reflecting the fact that it is always good for a gene to create cooperation 
within the close family, or, in other words, with itself. If that kind of
cooperation were already the rule, our newly created mutant population
would also be identified as family by its close relatives and the result
would change to $G=b$, or no improvement at all, but no worsening, as it
was obvious, as everybody is cooperating with everybody, regardless of
their genetic codes.

When we have only leaders ($p=1$), we obtain $G=a$, as nobody is 
cooperating. In a population where nobody cooperates, that would again
mean no difference at all. If cooperation with your family is already
the common strategy, we get $G = b( 1-\epsilon ) + a \epsilon$, which is
actually worse than the result for the non-mutants. If you have only 
leaders and no servants, the strategy is useless, as expected.

For the appearance of some leaders to bring some advantage to the only
servants case, we must see that $G$ increases as $p$ moves away from $0$,
or, in other words,

\begin{equation}
\left ( \frac{\partial G}{\partial p} \right )_{p=0}  > 0 \nonumber
\end{equation}

\noindent
which means $c+d>2b$. Therefore, we see that the strategy to acquire 
leaders only pays off when the average of $c$ and $d$ is higher than the
result one gets from the cooperative strategy, what makes sense. When we
have a Prisoner Dilemma where the parameters obey the $c+d>2b$ relation,
the appearance of a gene with two phenotypes, servants and leaders, is
possible. A population of non-cooperative individuals has no barrier
against such an invasion.

Let's now turn to the problem of invading a population where cooperation
already happens. We have seen that, in this case, an invasion by a
genetic code having only leaders can not happen as the result for such 
genes is actually worse than for those genes that do cooperate within 
their families. In this case, it is just a good idea to improve the
mutant strategy. If a leader can recognize who in his close family is
also a leader (using any kind of cues, like smell or body posture, the
details of this recognition are not important to the discussion) and who
is a servant, he might decide not to cooperate only with the servants
and cooperate with the other leaders inside his own family. As that
changes his results from the interaction with the other leaders from $a$ 
to $b$, that is actually an improved version that could easily replace
the strategy we were discussing so far in all the events where that
strategy was a winner. It is always useful for the leaders to cooperate
inside their family.

In this case, we see have that

\begin{eqnarray}\label{gain2}
G_{l} & = & \left ( 1-\epsilon \right ) b + \epsilon \left [ pb+\left ( 
1-p \right) c \right ] \nonumber \\
G_{s} & = & \left ( 1-\epsilon \right ) b + \epsilon \left [ pd+\left ( 
1-p \right) b \right ] \nonumber
\end{eqnarray}

\noindent
and

\begin{equation}
G = b \left ( 1-\epsilon \right ) + \epsilon S \nonumber
\end{equation}

\noindent
where

\begin{equation}
S= \left [ p^2 \left ( 2b-c-d \right ) + p \left ( c+d-2b \right )
+b \right ] \nonumber
\end{equation}

Here, we will have that $G(p=0)=G(p=1)=b$, that is, for both a
population with only leaders or only servants, the result is exactly the
same as for the non-mutant population. If once more we have $c+d>2b$, we
will have that, for any $p$ such that $0<p<1$ we have $G>b$. We have a
maximum for $G$ when $p=1/2$, therefore, we see that the population will
eventually mutate to a point where the numbers of leaders and servants are 
equal, as that point gives the highest gains.

Against a non-mutant population playing tit-for-tat, the outer strategy of 
the mutant gene just has to be tit-for-tat, as everybody. This way, the
gene makes it sure that they keep getting cooperation from the individuals
outside their family. In this case, both leaders and servants should keep
playing tit-for-tat, as they did before their gene mutated. However, inside
the family, we have seen that the division among leaders and servants, for
the right parameters in the Prisoner Dilemma, brings a better result than
simply cooperation. This way, the mutant gene can also invade an 
environment where everybody plays tit-for-tat and prosper there.

The equality in the number of leaders and servants means one very specific
thing. That, under a Prisoner Dilemma game, it is the best genetic reply to
have an equal number of individuals who increase their fitness as the number
of individuals who have their fitness decreased. When the population is
divided into these two groups, assymetric situations will arise as we will
disccuss in the next section and it is not clera right now if the $p=1/2$ 
would be always the best reply.

\section{Leaders and Servants in the General Two Players Case}

The result obtained in the last section was first developed as an 
alternative solution to the Prisoner Dilemma, but that is not its only
application. Let's examine now another type of game, one non-symmetric
game with complete information,  where a dominant strategy exists.
Non-symmetric games can happen among members of the same species easily,
specially if each member has different positions, as males and females, 
parents and children, or, if the leader-servant solution was
developed previously, among leaders and servants.
So far, it was believed that all genes should choose that strategy, as it
would be silliness to do otherwise, according with
traditional Evolutionary Game Theory. That is not true anymore if we 
introduce leaders and servants. 
Let's see how.

Two players, A and B, compete with 
each other, player A choosing the row and player B the line. Their payoffs 
are given by the pair

\begin{equation}\label{payoffa}
A = \left ( \begin{array}{cc}
		2 & 10\\
		4 & 3
\end{array}
\right ) \nonumber
\end{equation}

\begin{equation}\label{payoffb}
B = \left ( \begin{array}{cc}
		1 & 1\\
		2 & 3
\end{array}
\right ) \nonumber
\end{equation}

Here, it is clear that player B will chose the second line, as no matter
what A does, it is his best solution. Knowing this, either by rational
analysis, or simply observing that B always do that, A is limited to choose
between second line payoff and it will clearly picks the first column, so 
that the total result will be, for (A,B), (4,2). We have here an average
result of 3. However, were the two players to choose first line and second 
column, the average would go up to 5.5, thanks to very good result A would
be getting. Therefore, if these numbers represent, per example, the average 
number of
surviving offspring A and B will have, when facing this asymmetric situation,
it is a good strategy for a gene make its owners to choose the first line,
second column whenever faced with this decision. Again, we are supposing A 
and B can determine here, with reasonable success, that they share this same
gene, or in other words, that they are close relatives, something we will go 
on assuming in this paper.

Here, if the gene of player B changes, so that he will not cooperate for A to
get a better result, B will prosper initially, but, as his descendents form
their own family, where the components take always the dominant strategy, they
lose the advantage our mutant gene had and, as consequence, evolution works 
to take them out of the scenario.

The same dynamics can happen for a number of other games, including games with 
multiple Nash equilibria, where one of the equilibria is clearly more 
favorable to the gene than the other one.

A point that should be very clear is that being a leader has not 
necessarily anything at all to do with be the one who gives the orders or 
make the decisions. Being a leader, in this context, means just that other 
beings from your family will sacrifice their own fitness in order to 
increase yours. They might even be the ones responsible for the decision 
making and you nothing more than a reproductive machine. The point of view 
here is not a subjective one, but an evolutionary one, where all that 
matters is whether you will leave fertile children or not. Evolution is not 
concerned with issues like freedom or happiness, unless they would 
represent some change in your fitness.

\section{Flexibility, Specialization and Multi-Cellular Organisms}

An important genetic decision is on strategies to decide who would play 
the role of leader and who 
would be the servant. More specifically, it is interesting to ask if these 
positions sholud be perpetual. We can reason that it is not clear whether a 
leader should remain leader for his entire life as 
there it may happen that it is evolutive better for the roles to be traded. 
A gene that, once the decision about 
which role a individual would be playing was made, would mark the leaders as 
leaders for life as well as the servants  
would be less flexible and possibly less efficient than one that 
would allow a change of ranks. Flexibility seems a good idea, at first. The 
flexible 
gene has to develop a way to determine when it is convenient to change the 
hierarchical positions, based on the 
actual information the individual has access to, and, maybe, on the amount of 
resources already invested in the old leaders. In the extreme 
case, supposing a well tuned mechanism of change,  the change wouldn't be 
necessary and this strategy would not cause any 
changes. Being as efficient as the non-flexible variety. In all the other 
occasions, the flexible gene would use its flexibility to get improved 
chances. In human societies, we see new leaders taking the place of old 
ones. That may be this effect of changing the position of leaders and servants 
working, or simply the rise of a new 
generation, as the older individuals, due to age, have their fitness 
decreased and it is useful for the species to put other leaders in their 
places.

This way of doing things should lead to change in the individual 
characteristics, associated with a change in the hierarchical position.
This effect can be a good way to determine if the leaders are turning
into servants or if they are just been replaced by new leaders. If the old
leader, when demoted, loses its regal posture and seems weaker, as happens
with some of them, we are actually seem his position been reverted. The
opposite is hardly true. The only people who ever makeit into leader
positions seem to be those who were already leaders of smaller groups.
However, good fortune can change positions bringing power and/or money to
a servant. In these cases, the opposite change, from servant to leader, is 
to be expected. 

On the other hand, very rigid structures, where no changes at all are 
possible, allow a better specialization. A 
servant who is always a servant can specialize his work and get a better 
result for his family and genetic code, like in an ant colony. The gene could
mutate to include an instruction that, if the individual is turned into a 
servant, his development would happen according to some blueprint; if he is a
leader, his development to adulthood is changed to fit his different 
hierarchical position. 

If specialization wins, this can lead to the appearance of large social 
structures and multicelular organisms. It has already been pointed that, 
the fact that all the beings in a community evolving towards a single 
organism share the same genetic code, makes cooperation a good idea, as that 
gene has its survival chances improved. Here, not only cooperation happens,
but some individuals use strategies athat decreases their fitness, in order
to improve their leaders' fitness. In the limit situation, the servants might
give up completely on having descendents of their own. The leaders make up
for this by having descendents who are servants and others who are leaders,
keeping the population with the same distribution as before. If the servants
start contributing with very few descendants to the next generations, they 
soon stop having any evolutionary utility but increasing their leaders 
fitness.
 
That way, unicellular beings may have found, at some stage of their evolutions, 
that creating infertile descendents was a very good idea.  Those beings would
be programmed with some specific task, like protection,
improved food gathering or anything else, whose function was to serve the
ones who would be responsible for the procriation. That 
can lead to cells who were expert in their tasks, unable to survive on 
their on, but that contribute to the whole. In this sense, we could understand
that the whole body of cells that make up a multicellular being are all slaves
to the sexual cells, who are the ones that really make copies of themselves and 
of everybody else, of course.
The genetic code survives and alters itself exactly by that reproduction. And,
of course, the sexual cells, or leaders in the framework we developed above, 
want to create new servants (actually, the other parts of the body) or else
they would fail in their task to get reproduced. 

When the dynamics between flexibility and specialization comes to a stable 
solution, we have the creation of a new
kind of being, a multicelular being, with no inner struggles about who should
get the better fitness. The servant cell fitness can decrease to the point 
where
they only reproduce to repair damaged tissues, and they don't even try to get
an evolutionary advantage for themselves.

The same process can work with multi-cellular entities and we will have 
structures like ant colonies, that are actually one only being, from an
evolutionary point of view. The question of why some ants work only as 
soldiers, never 
getting reproduced, living just to be sacrificed, actually does not make
sense from an evolutionary point of view. Those soldiers do not reproduce, so
their fitness is not an issue when determining any type of evolutionary
success. During the process they when they became servants, their fitness was
decreasing and becoming each time less important for the species as a whole.
Therefore, it was just expected that they have turned into expendable 
servants.

Both forces, for flexibility and for specialization, will compete and the 
most apropriate structure for the situation will win.
It is quite possible, from what we see in nature, that the rigidity has a 
tendency to win, 
forming new individuals, superorganisms composed of the smaller organisms 
working together as one. It has won in the multi-cellular case, we see very 
few examples of colonies of microorganisms working together, with fertile 
leaders and
servants present. While superorganisms consisting of very specialized servants
cells working to improve the fitness of a few cells are very common and are all
around us, like ourselves, or every animal or plant that we find. Specialization 
has won among the social insects. Among us, the struggle between the two forces 
still exists, but this may be just a transient phase. Only the future can tell.

\section{Some Comments and Possible Applications}

In this section  will allow myself to speculate on the possible 
explanantions to natural phenomena provided by this result.
I will ask the reader to forgive me if she thinks
I have gone too far, my only defense is that I do believe that what I will be
saying here is, if not a true description of Nature, at least, very 
reasonable and likely.

A first important warning is that one should be very careful when trying to
identify situations where leaders and servants have appeared. This because
these terms were used here in a very strict sense. Servants are not individuals 
who obey the leaders, but individuals who take strategies that decrease their 
fitness in order to increase their leader's fitness. In this sense, there
are situations where leaders and servants might, according to our definition
be identified in the reversed positions than leaders and servants in every
day language.

We know that it can be a good decision for a gene to spend some of its 
carriers to further 
the survivability of other carriers. We have seen how this mechanism works,
how it can be the right evolutionary thing to do to sacrifice some 
individuals for a greater "good". It can not be stressed enough that this
does not mean any of us must agree with this genetic moral. The point
is just that moving resources from the servants to the leaders, when facing a
problem with the appropriate parameters, can be a winner evolutionary
strategy. Therefore, living beings can have inside them some kind of
structure that makes this type of slavery not only possible, but even 
desirable by the servants. This 
slavery is not necessarily of the type we humans are used to recognize as
such, of course.

All this can lead us to some speculation on how this genetic strategy can be
actually seen working in nature and in ourselves.
It seems to me that much of our society could have been built around such 
idea. In the old days, our communities were much smaller and people from
the same tribe as we were probably our relatives. Therefore, it was a good
genetic strategy to divide the human society in leaders and servants and to 
base our decision on whether someone belonged or not to our family on the
fact that they lived in the same tribe we did.

We have inherent abilities to recognize hierarchy and our position in it 
and we expect to be treated 
accordingly. When a person is low in the hierarchy (a servant) it accepts 
abuses it wouldn't accept if its position were higher (a leader). We have 
always heard that power corrupts. Well, when you are a leader, you should
expect to be treated like one, that is part of the strategy. Per example,
when a family has two children, the parents may decide to invest more in 
one of them and simply ignore the other one, if that would make the family
fitness higher. The servant child will be expected to work and actually help
feeding and fulfilling the needs of the leader, so that the leader child will
have even better chances. More food, protection, more access to suitable 
members of the opposite sex.

When detecting leaders, human beings seem to use other clues than posture to
determine 
the hierarchical position of someone. More specifically, we use clothing to 
represent our position and this seems to be a trace common to many 
cultures. Therefore, it is not so strange that people worry far more than 
would seem reasonable about fashion and tend to prefer more clothing that 
is more expensive just because it is more expensive and not because it is 
better or more difficult to make.  If wearing clothes is something people 
will use to determine if we are leaders or servants and decide whether to 
cooperate or not with us, it is not strange at all that people, since a 
very young age, seem so irrationally attracted to clothes. That can be not 
just a way to show to which group you belong but also a strategy to be seen 
as a leader.

The teenage struggles to be accepted by their peers become now even more 
desperate. If all of them would play tit-for-tat in their social 
relationships, they would not have many problems among them. But that's not 
the point of the game, the point is to establish oneself as leader, as 
someone the servants must cooperate with and not expect cooperation in 
return. Therefore, it is to be expected problems at some point before the 
adult life, some period when, if left to themselves, humans will fight each 
other, physically or verbally, trying by all means make themselves leaders 
and those they don't like servants. And we all know what happens when one 
is a teenager.

The same effect should be expected around organizations and this could be
checked. Organizations seen as leaders, with prestige, should be able to 
atract workers with smaller wages for the same task than declining 
organizations, as people would want to be seen as leaders. 
It is also very interesting to see how many organizations 
make it a very clear point to tell people they are servants as soon as they 
are accepted as part of the organization. Universities, with the treatment 
freshmen get from their just a little older fellows, military 
organizations, the examples are many. Make the man know he is a servant and 
he will cooperate, even when the organization, or its leaders, do not.

Other area where this effect can happen is in the field of ideas and in 
politics. Some ideas are know to send its followers to paths that actually
make their fitness lower, but the idea still flourish. A priest who is told he
can't have sexual relationships, much less, children; a terrorist who dies for 
the cause, all have their fitness terribly diminished by these strategies. 
However, as we do have the ability to become servants, they can take these 
paths. In this case, their paths might serve somehow to further the goals of 
the ideas they are slaved to. But they can also serve to increase the fitness
of the other member of the man's tribe, be it a rligious community or a
military party.

Our leaders do think about the life of their soldiers as something expendable
in order to get to their objectives. Objectives that make them more powerful,
possibly allowing their genes, or the dominant genes in their country, to 
dominate and spread over greater areas, therefore, increasing the fitness of
their own genetic code. We have a great sense of hierarchy, we are constantly
gauging our position in the social hierarchy and we tend to behave according to 
that. Our judgement may be different from that accepted by the official 
society, as in the case of a mob leader, but, even there, there is a hierarchy 
to be respected, it is always clear who the leaders are. We instinctively tell 
in our body stance, in the clothes we wear, where in that hierarchy we are and 
that knowledge is used by everybody to decide how to behave around us.

In the old days, we would just normally meet and call friends people who 
belonged to our tribe and who were probably our relatives. Working with them 
and playing leaders and servants then was a good choice. Nowadays we carry the
same genes, but most people we meet are not family. Our systems to decide who
we might expect to share a servant-leader relationship do not work as well
anymore and can be exploited by our bosses, religious political leaders. It is
no surprise to me that one of the most common practices in this fields is to
make you feel like you are part of a family. Because people not only
cooperate with their families, they also accept servant roles in that position
and make sacrifices they wouldn't do otherwise.

This has other possible applications. Police officers are notoriously 
known, to a greater or smaller extent, in several parts of the world and/or 
periods of history, for abusing their functions.  Political leaders 
have done it even worse, some considering that nothing was more important 
than their personal desire. In both cases, we see people in positions that 
give them power, using this power in a non-cooperative way, expecting that 
the servants would obey them. The more power given to a person, the more 
she requires from her followers, and the more corrupt she becomes. What 
could be happening here is simply the fact that this person, who learned 
she is the supreme leader, knows instinctively that she must not cooperate 
with the servants. That can be the best genetic strategy. Therefore, 
it is our own genes that make us corrupt as we grow 
more and more powerful.

It is probaably more than a coincidence that, as culture changed and we 
start to tell our 
leaders and police men that they are not the real leaders, but servants of 
the rest of society, their behavior changes towards a more cooperative one. 
They still have power and the desire to use it and become leaders but, as 
the society do not treat them as leaders, the problems with abuses 
decreases.

Turning our attention back to psychological problems, it is interesting 
to ask how we notice and classify who is a leader and who is a servant. 
Posture 
is a good guess as well as tips we can give with our behavior. Other subtle 
ways can be working as well and some research to identify them could lead 
to interesting discoveries.

This dynamics can be used also to explain the apparently irrational success 
of self-help 
literature and movements. If you can convince people to behave as leaders, 
and the world will react to that by treating them that way, it is very 
likely they their lives will get better, as more people start cooperating 
with them. The down side to it is that their improvement happens with 
servants sacrificing themselves for the new leaders; this way, they are not 
really making the world a better place, just exploring it and making sure 
they get their best share of it.

Aggression inside families can be a mean to determine leadership. As we 
have seen, it is exactly inside a family that the 
leader-servant strategy is expected to work best, all other cases been 
just weaker variations, based on erroneous instinctive methods to determine 
family. What we have is that violence can be a mean for the leader to 
establish his dominance. Servants can't be violent with their leaders, as 
that would make their leaders fitness worse, but, under the right 
circumstances, leaders could be violent 
and the servants would still serve them. 

When someone makes 
himself a leader using brute force, it would be reasonable to expect that 
the other person would, if he were acting according to a dominant strategy 
for his personal interests, do something. Taking the problem to 
the police, changing relationships, fighting back, all are better personnal 
strategies than just accepting the violence. However, when servants 
recognize their position as servants, their instincts are expected to work 
towards keeping them in that position. Therefore, it should be very hard 
to convince a victim of aggression to act against her aggressor, when this
happens inside a family.

In the economic arena, this type of dynamics is also known to happen. 
Conglomerates can force their component companies to take bad individual
decisions, specially when dealing with companies from the same group, 
so that, in the whole, the conglomerate will profit more. As long as the 
group ends up better, they would never care about making one of their
companies worse.

\section{Conclusion}

We have seen that when genes are playing their games, it is not always in their
best interest to play it so that the individuals playing their parts would 
always the problem answer with a dominant strategy. There are situations when
it is genetically preferred to have individuals making decisions against their 
best interests, so that the whole survivability of the gene gets increased.
As long as the servants have a mean to adequately determine who are the leaders
with the same genes as them, they can use that information to increase the 
leader's fitness, even when it decreases their own. This solution is not 
evolutionary unstable, if the individuals adopting it belong to the same
family.

The major consequences of this is that not only cooperation becomes more likely,
helping to explain how it is so common in our word, but also that there will be 
times, even in non-symmetric games, when a dominant strategy for the gene 
carriers is not the best answer, from the gene point of view.

\end{document}